\documentclass[twocolumn,showpacs,amsmath,amssymb]{revtex4}

\usepackage{graphicx}
\usepackage{dcolumn}
\newcommand{\nn}{\nonumber}
\newcommand{\beq}{\begin{equation}}
\newcommand{\eeq}{\end{equation}}
\newcommand{\be}{\begin{eqnarray}}
\newcommand{\ee}{\end{eqnarray}}

\def\+{\dagger}
\def\la{\langle}
\def\ra{\rangle}
\def\<{\langle}
\def\>{\rangle}

\begin{document}


\author{James Charbonneau}
 \email{james@physics.ubc.ca}
\author{Ariel Zhitnitsky}
 \email{arz@physics.ubc.ca}
\affiliation{
Department of Physics and Astronomy,
University of British Columbia,
Vancouver, BC, Canada, V6T 1Z1
}

\title{A Novel Mechanism for Type-I Superconductivity in Neutron Stars}

\date{\today}

\begin{abstract}
We suggest a mechanism that may resolve a conflict between the precession of a neutron star and the widely accepted idea that protons in the bulk of the neutron star form a type-II superconductor.  We will show that if there is a persistent, non-dissipating current running along the magnetic flux tubes the force between magnetic flux tubes may be attractive, resulting in a type-I, rather than a type-II, superconductor.  If this is the case, the conflict between the observed precession and the canonical estimation of the Landau-Ginzburg parameter $\kappa > 1/\sqrt{2}$ (which suggests type-II behaviour) will automatically be resolved.  We calculate the interaction between two vortices, each carrying a current $j$, and demonstrate that when $j > {\hbar c\over 2 q \lambda}$, where $q$ is the charge of the Cooper pair and $\lambda$ is the Meissner penetration depth, a superconductor is always type-I, even when the cannonical Landau-Ginzburg parameter $\kappa$ indicates type-II behaviour.  If this condition is met, the magnetic field is completely expelled from the superconducting regions of the neutron star.  This leads to the formation of the so called intermediate state, where alternating domains of superconducting matter and normal matter coexist.  We further argue that even when the induced  current is small $j < {\hbar c\over 2 q \lambda}$ the vortex Abrikosov lattice will nevertheless be destroyed due to the helical instability studied previously in many condensed matter systems. This would also resolve the apparent contradiction with the precession of the neutron stars.  We also discuss some instances where anomalous induced currents may play a crucial role, such as in neutron star kicks, pulsar glitches, the toroidal magnetic field and the magnetic helicity.  
\end{abstract}



\maketitle

\section{Introduction}
This paper is motivated by calculations \cite{Link:2003} that show a conflict between the observed  magnitude ($\sim 3^\circ$) and frequency ($\sim 1$ per year) of neutron star's precession \cite{Stairs:2000} and the widely accepted idea that protons in the bulk of the neutron star form a type-II superconductor.  We begin with a review of this contradiction.

\subsection{Precession and Superconductivity}
In the generally accepted picture, the interior of a neutron star contains neutrons and a small number of protons and electrons.  A compilation of neutron and proton scattering data \cite{Sedrakian:2006} implies that the extremely cold ($10^8$ K), dense ($10^{15}\text{ g/cm}^3$) nature of the neutron star should cause the neutrons to form $^3 P_2$ Cooper pairs and become a superfluid and the protons to form $^1 S_0$ Cooper pairs and become a superconductor.   

A rotating superfluid cannot form a solid body to carry circulation but instead forms vortices of quantized circulation that run in the direction of the angular velocity \cite{Feynman:1955}.   It is also generally accepted that the protons form a type-II superconductor, which means that it supports a stable lattice of magnetic flux tubes in  the presence of a magnetic field   \cite{Landau:9}.  The rotation of the neutron star ($1-10^3$ Hz) and the presence of the enormous magnetic field ($\sim10^{12}$ G) are sufficient for both superfluid and superconducting vortices to form inside the neutron star.  It should be noted that when a superconductor is placed in a rotating container it co-rotates with the container at the expense of a small current known as the London current \cite{Sauls:1989}.   

Vortices formed in a superfluid will always repel and form a lattice.  Vortices formed in a superconductor can either attract, in which case the magnetic field is expelled from a superconductor, or they can repel each other and form a triangular lattice.  These two behaviours label the superconductor.  If the vortices attract, the superconductor is called type-I and if they repel it is called type-II.  The distinction between the two types of superconductivity will be very important in this paper.

The standard way to determine the type of superconductivity is to calculate the value of the Landau-Ginzburg parameter \cite{Landau:9}, $\kappa = {\lambda \over \xi}$, where $\lambda$ is the Meissner penetration depth and $\xi$ is the coherence length of the superconductor.  If $\kappa > 1/\sqrt{2}$ then the superconductor is type-II, otherwise it is type-I.  This parameter is weakly dependent on density.  Typically for nuclear matter in a neutron star, $\lambda \sim 80$ fm and $\xi \sim 30$ fm \cite{Link:2003}, which means that $\kappa \sim 2.6$.  At higher densities it is possible to have a Landau-Ginzburg parameter that indicates type-I behaviour \cite{Sedrakian:1997}, but only in the densest part of the core.  There still exist large regions of the neutron star where type-II behaviour is predicted. 

In a neutron star both condensates are subjected to angular momentum and magnetic flux and form lattices, but the proton vortices ($\sim 10^{24} \text{ per m}^2$) are much more numerous than the neutron vortices ($\sim 10^{10} \text{ per m}^2$) and are tangled around them.  It is the formation of these lattices that cause the contradiction with the precession of the neutron star. 
  
The existence of precession means that the superfluid neutron vortices no longer form along the rotational axis of the star, but along the axis that is the sum of the precession and angular momentum vectors.  When the star precesses the vortices now move with respect to the rotation of the star and, in turn, with respect to the proton vortices which are entangling them.  If the precession is large enough one of two things must happen; either the neutron vortices move with the proton vortices or they pass through each other.  

Requiring the neutron and proton vortices to move together places severe restrictions on the precession.  This is the case of "perfect pinning" that was discussed in \cite{Shaham:1977} using macroscopic dynamics.  We will follow the arguments in \cite{Link:2003} as we are specifically interested in proton vortices as a mechanism for pinning.  Because the core of the star is superconducting, the proton vortices, which carry magnetic flux, are resistant to being moved \cite{Baym} and thus the neutron vortices are restricted to move slowly.  This means that the neutron vortices are pinned to the rotation of the protons and thus are pinned to the rotation of the crust.  If this pinning is present the neutron star can only precess at very high frequencies.  If the star is to precess more slowly at large amplitudes then it is necessary for the neutron vortices to pass through the proton vortices.

The case of "imperfect pinning" was first discussed in \cite{Sedrakian:1999} using the concept of vortex drag. Microscopically, this drag is created by large numbers of neutron vortices passing through proton vortices \cite{Link:2003}.  This creates a number of excitations and is a highly dissipative process.  Both methods find that the precession is highly damped and that there are no modes of large, persistant precession.   
Based on these estimates it is concluded that given the observed precession neutron vortices and proton flux tubes cannot coexist in the star \cite{Link:2003}.  Either the star's magnetic field does not penetrate any part of the core that is a type-II superconductor or that at least one of the hadronic fluids is not superfluid.  Based on pairing calculations that predict neutron and proton superfluids coexist in the outer core the latter is very unlikely, so we will look to the former.  

If the core is a type-I superconductor, the magnetic flux could exist in macroscopic regions of normal matter that surround superconducting regions known as an intermediate state.  The magnetic flux would not form proton vortices and there would be nothing to impede the movement of the neutron vortices, thus allowing the star to precess with a long period. 

For proton vortices to not form a lattice, a mechanism must be present to make the interaction between them attractive (type-I behaviour).  If it does so even when the cannonical Landau-Ginzburg parameter suggests that the superconductor is type-II then the inconsistency described above will be resolved. 

\subsection{A New Mechanism for Type-I Superconductivity}
In this work we suggest a mechanism that leaves the Landau-Ginzburg parameter unchanged, but causes the system to behave quite differently than in the standard picture.  To be precise, we will show that even when $\kappa > {1\over \sqrt{2}}$ the system prefers the intermediate state and that the apparent contradiction between $\kappa \sim 2.6$ and the observed precession is avoided. 

This is achieved if the system supports a persistent, non-dissipating current running along the core of a vortex. Such topological currents appear in many systems as the consequence of a quantum anomaly.  It is well known that anomalies, and the topological currents they induce, have important and non-trivial implications; the electromagnetic decay of neutral pions $\pi^0\rightarrow 2\gamma$ is a textbook example.  

As we will discuss later, analogous topological currents have even been observed in some condensed matter systems.  However, with a few exceptions, the analysis of quantum  anomalies has not received attention in the literature devoted to dense matter systems in general, and neutron stars in particular. In section II  we discuss in detail why and when such topological currents may arise in high density systems.

If such currents are induced, they drastically change the  behaviour of the system.  Normally the interaction between superconducting vortices has two terms: an attractive force which comes from the order field and a repulsive force which comes from the gauge field.  In the presence of an induced current, a third, attractive force will appear and, if the current has sufficient magnitude, the system will behave as a type-I superconductor.  This situation will be discussed in sections II and IV.   
 
\subsection{Relation to Previous Work}
The same  problem has been  discussed previously in \cite{Sedrakian:2005},  \cite{Buckley:2003zf}. In particular, \cite{Sedrakian:2005} has shown that the existence of a type-I superconductor in a neutron star resolves the conflict addressed in \cite{Link:2003}.  Reference \cite{Sedrakian:2005} starts by assuming that the equilibrium structure of a type-I superconductor contains alternating superconducting and normal domains.  By use of a hydrodynamic restriction based on the moment of inertia of the crust and the moment of inertia of the superfluid it is shown that the alternating domain structure seen in type-I superconductors will always allow for undamped precession.  

This result can be understood in a different  way.  The arguments of reference \cite{Link:2003} relied on the proton vortices tangling around the neutron vortices.  In a domain structure there is more room for the neutron vortices to move unhindered by the proton vortices allowing for large amplitude, high frequency precession.

However, the  calculation in reference \cite{Sedrakian:2005} starts by assuming that type-I superconductivity already exists in neutron stars.  It does not investigate how a type-I superconductor could arise and it is not obvious how to reconcile this with standard arguments that suggest $\kappa \sim  2.6$, which ambiguously implies that the superconductor is type-II in a finite volume of the neutron star.

In reference \cite{Buckley:2003zf} a mechanism has been suggested that potentially resolves the conflict between the standard estimation of $\kappa$ and type-I behaviour.  The mechanism is based on the idea that proton and neutron Cooper pairs have almost identical interactions, akin to their underlying isotopical symmetry, even though the Fermi momenta and densities for protons $\mu_\text{p}$ and neutrons $\mu_\text{n}$ are vastly different. A small difference in interactions was modeled by a small effective asymmetry parameter $\xi \ll 1$.  In this case it has been shown that the type of superconductivity is not governed by the canonical Landau-Ginzburg parameter $\kappa$, but instead, by an effective Landau-Ginzburg parameter $\widetilde{\kappa} = \xi\kappa$.  This effective parameter can be small,  $\widetilde{\kappa} < {1/\sqrt{2}}$ leading to type-I superconductivity while keeping canonical Landau-Ginzburg parameter large, $\kappa > {1/ \sqrt{2}}$~ \footnote{The basic assumption of ref.  \cite{Buckley:2003zf} has been criticized in \cite{Alford:2005ku}.  It is not the goal to discuss the approach developed in \cite{Alford:2005ku} in the present paper, but a short remark is warrented.  If one uses the technique from \cite{Alford:2005ku} for QCD(N$_\text{c}$=2) (which can be solved exactly for $\mu_i \ll \Lambda_\text{QCD}$) one should anticipate similar results, namely that vastly different densities for different flavours would lead to very different scattering lengths for different flavours.  The exact solution for QCD(N$_\text{c}$=2) teaches us differently: that different flavours interact in the same way as a result of the underlying  flavour symmetry despite the possiblility of vastly different densities for different flavours.  More studies are required before the basic assumption of ref. \cite{Buckley:2003zf} can be shown to be incorrect.}.

\section{Persistent Non-Dissipating Topological Currents and Vortices} 
Though the idea of non-dissipating topological current in vortices was considered long ago \cite{Witten:eb} in the context of cosmic strings, we are more interested in the recent developments of similar phenomena in high density QCD \cite{Son:2004tq, Metlitski:2005qz, Metlitski:2005pr, Newman:2005as}  and condensed matter systems \cite{volovik1, volovik2, Alekseev:1998ds}.  The most important result can be formulated  as follows.  

Consider QCD with non-vanishing chemical potentials, $\mu_L$ and $\mu_R$, which correspond to two reservoirs of particles with different chirality.  Due to the chiral anomaly a number of interesting macroscopic phenomena could occur: induced non-dissipating vector currents on an axial vortices, induced axial currents on a vector vortices, magnetization of the axial domain wall, induced angular momentum by vortex loops, to name just a few.
 
Presently we are interested in the phenomena where currents are induced in the background of an external magnetic field.  To be precise, in the chiral limit and zero temperature $(m_q=0, T=0)$, each fermion species $q$ makes an additive contribution to the vacuum expectation values for the axial and vector currents,  
\be
\label{AV}
\la \int_S \mathbf{j^A}\cdot d\mathbf{S} \ra= {e (\mu_R+\mu_L) \over 4 \pi^2}\Phi\,, \\
\la \int_S \mathbf{j^V}\cdot d\mathbf{S} \ra= {e (\mu_L-\mu_R) \over 4 \pi^2}\Phi\,.\nn
\ee
where $\Phi= \int d^2x_{\perp}B^z(x_{\perp})$ is the total magnetic flux through the cross section $S$, and the fermion current densities are defined as $j^A=\bar{q}\gamma^3\gamma^5q, ~ j^V=\bar{q}\gamma^3q $.  A few comments are in order.  Formula \eqref{AV} has a universal nature, as it originates from the fundamental quantum anomaly and it is not sensitive to whether the magnetic field is localized inside of vortex or uniformly distributed over a large area $S$.  Also, corrections due to non-zero fermion mass and temperature can be explicitly calculated, but will not be discussed in the present paper. 
 
It is worth noting that similar non-dissipating currents have been discussed in condensed matter literature \cite{volovik1, volovik2, Alekseev:1998ds}.   In particular, the expression for anomalous supercurrent has been derived for $^3$He-A system based on the chiral anomaly (see eq.(5.35) and  Fig.23 in ref.\cite{volovik1}).  Some suggestions of how these effects can be experimentally tested were also presented in ref.\cite{volovik1}.  An important feature of the $^3$He-A system is the existence of the anomalous chiral symmetry, which is not present in $^4$He.  Therefore this phenomenon exists in $^3$He-A but not in $^4$He.  

Anomalous supercurrent may also exist in high $T_c$ superconductors with $d$-wave paring  or in a graphene system when the relevant degrees of freedom satisfy the massless Dirac equation, with a velocity of order one hundredth  that of light. In these cases, the chiral symmetry is obviously present and there is a good chance that non-dissipating, topological currents may exist.
  
For now we will restrict our discussion to QCD.  Note that in case of equal chemical potentials, $\mu_R=\mu_L=\mu$, the vector current is not induced due to the exact cancellation between left handed and right handed fermions.  This would be exactly the case in normal nuclear matter when $\pi$ meson condensation does not occur or in color superconducting phases, such as CFL at asymptotically large chemical potentials, where Goldstone modes do not condense (see recent review \cite{Alford:2006fw} and references therein on color superconductivity).  

However, it is known that Goldstone modes are likely to condense in nuclear matter \cite{Kaplan:1986yq,Takahashi:2001jq} and will definitely condense in color superconducting phases for intermediate chemical potential \cite{Schafer:2000ew}, \cite{Kaplan:2001qk}, \cite{Kryjevski:2004cw}, \cite{Alford:2006fw}. In the cases of neutral $\pi^0$ condensation in nuclear matter \cite{Takahashi:2001jq} and $\eta$ condensation in the color superconducting phase \cite{Kryjevski:2004cw} the vector current will definitly be induced, as shown by the simple argument below.  

The condensation of charged Goldstone mesons $(\pi^{\pm}, K )$ in the system is much more complicated, but we can expect that the condensation of the charged pseudo-scalar Goldstone mesons in the system will lead to different densities for $L$ and $R$ species (and corresponding to different  effective chemical potentials for $R$ and $L$ modes), in which case the vector current will be also induced.
 
It is not the main goal of this paper to describe all possible phases where axial density (and consequently, the vector current) can be induced.  Rather, we want to give a simple argument demonstrating why the condensation of a pseudo-scalar Goldstone mode would necessarily lead to a difference in densities for left handed and right handed species.  

To simplify arguments, we consider a model with  just a single flavor. Let us assume that we are in a phase where the baryon density is non-zero and a neutral $\eta $ Goldstone mode   is condensed. This implies that our ground state can be understood as a coherent superposition of an infinitely large number of the Goldstone $\eta$ mesons.  We expect that the ground state of the system is not disturbed by adding one extra $\eta$ meson into the system.  On the other hand, we can relate the matrix element with an extra $\eta$ meson to the matrix element without the $\eta$ using the standard PCAC technique, $\< A| O|A \eta\>\sim i\< A| [O, Q^5]|A \>$.  In the present case the coefficient of proportionality would not be precisely $1/F$ (where $F$ is the Goldstone coupling constant) because our Goldstones are in the $\<\eta\> $ condensed phase rather than in a trivial vacuum.  

Taking $|A\>$ to be the ground state and $O$ the baryon density operator, one can immediately see that if baryon density does not vanish in the ground state, then the axial density $\<  [O, Q^5] \>$ will not vanish also\footnote{ It is interesting to note that while the expectation values of different operators in dense matter (such as baryon density, diquark condensate, chiral condensate, pion condensate, etc) have been calculated a number of times, in a number of models (see e.g. review \cite{Alford:2006fw}), the calculation of axial density has not received much attention.  In fact,  while axial density is generically non-zero when a Goldstone mode is condensed, as argued above, it has been calculated only recently in $QCD (N_c=2)$ and in $QCD (N_c=3)$ with isospin chemical potential $\mu_I\neq 0$ \cite{Metlitski:2005db}.  In these cases the exact expressions can be obtained for the axial density in the regime $\mu_I\ll \Lambda_{QCD}$.  An important lesson that \cite{Metlitski:2005db} teaches us is that the axial density does not vanish in spite of the fact that the corresponding axial chemical potential was not explicitly introduced.  Rather, it was generated dynamically.  The presence of the axial density unambiguously implies that $R$ and $L$ handed modes have different densities, and therefore, the vector current \eqref{AV} will be induced.}.  This implies that densities for left handed and right handed species are different and therefore a vector current will be induced. We assume this to be the case in what follows. 

We should note that the condensation of the pseudo scalar Goldstone mode is not the only mechanism capable to produce the asymmetry between $R$ and $L$ modes; any P parity violating processes can do the same job. The crucial point here is not the ability to produce the asymmetry (which is a common phenomenon in neutron stars due to the neutrino emission), but the ability of the non-dissipating persistent currents to keep this asymmetry through the entire volume of the star and deliver it to the surface of the star.  The phenomenological significance of this is argued in section V. 
 
As all quarks have non-zero electromagnetic charges, once a vector current is induced, an electromagnetic current will also be induced.  Our next step is to derive the interaction between two superconducting vortices where an induced electromagnetic current is present in their cores.  Our ultimate goal is to understand (at least qualitatively) the changes which will occur in the system due to these induced currents.  This is the subject of the next two sections.  
  
\section{Structure of a current carrying vortex}
In formulating the problem we assume that the superconductor is due to non-relativistic proton Cooper pairing though similar results are also valid for phases where a relatevistic field theory should be used.  It does not change the qualatative picture described below.  We start with the two dimensional Landau-Ginzburg free energy with term added to model a current source $\mathbf{j}$.  As discussed earlier, this current is an induced persistent electromagnetic current and it couples to the gauge field naturally.  Dependance along the third direction is neglected so $E$ measures free energy per unit length,
\begin{eqnarray}
E &=& \int{d^2x}\left\{{\hbar^{2} \over 2m}\left|\left(\nabla - {iq\mathbf{A}\left(\mathbf{x}\right) \over \hbar{c}}\right)\psi\left(\mathbf{x}\right)\right|^{2} - {\mu_\text{b}}\left|\psi\left(\mathbf{x}\right)\right|^{2} \right. \nonumber \\ &&  \left. + {a \over 2} \left|\psi\left(\mathbf{x}\right)\right|^{4} + {1 \over 8\pi}\left(\nabla\times\mathbf{A}\left(\mathbf{x}\right)\right)^{2}+{1 \over c}\mathbf{j} \cdot\mathbf{A}\right\}\,,
\end{eqnarray}
where $\mu_\text{b}$ is the chemical potential of the Cooper pairs and $a$ is related to the scattering length $l$, $a = {4\pi \hbar^2 l\over m}$. The $\psi$ field describes only the superconducting component of the protons.  Remember that the fundamental particle here is the proton Cooper pair whose mass is actually $m = 2 m_p$ and its charge is $q = 2|e|$.  The current source that was added will be used to model the current flowing along the core of the vortex.  As discussed earlier, this current can be treated as an external   electromagnetic current.    

The Landau-Ginzburg free energy without the extra term for the current source has a symmetry under the following gauge transformation,
\begin{eqnarray}
\label{symmetry}
\mathbf{A}\left(\mathbf{x}\right) &\rightarrow& \mathbf{A}\left(\mathbf{x}\right) + \nabla \varphi\left(\mathbf{x}\right)\,, \nonumber \\
\psi\left(\mathbf{x}\right) &\rightarrow& e^{{iq \over \hbar{c}}\varphi\left(\mathbf{x}\right)}\psi\left(\mathbf{x}\right)\,.
\end{eqnarray}
We assume that the external current $\mathbf{j} $ is conserved $\nabla\cdot\mathbf{j} = 0 $ in which case (\ref{symmetry}) obviously remains a symmetry.

We will choose a form for our current source, $\mathbf{j} = j \delta^2(\boldsymbol{x})\boldsymbol{\hat{z}}$, which models a current at $\mathbf{r}$ traveling in the $\hat{z}$-direction.

The free energy can be minimized to yield the the equations of motion.  Minimizing with respect to the vector potential $\mathbf{A}$ yields a Maxwell equation for our system, 
\begin{equation}
\label{maxwell}
{1 \over 4\pi}\left(\nabla^{2}\mathbf{A} - \nabla\left(\nabla\cdot\mathbf{A}\right)\right) = -{\hbar{q} \over m} \mathbf{j}_\text{Noether} - {1\over c}\mathbf{j}\,.
\end{equation}
The right hand side is written in terms of the Noether current 
\begin{equation}
\label{current} 
\mathbf{j}_\text{Noether} =  {1 \over 2i}\left(\psi^\dagger\nabla\psi - \psi\nabla\psi^\dagger\right) - {q \over
\hbar{c}}\mathbf{A}\left|\psi\right|^{2}\,.
\end{equation}
Minimizing with respect to the order field $\psi$ gives
\begin{equation}
\label{order}
{\hbar^{2} \over 2m}\left(\nabla - {iq \over \hbar{c}}\mathbf{A}\right)^{2}\psi = a\left|\psi\right|^{2}\psi - \mu_\text{b}\psi\,.
\end{equation}

For determining the structure of the vortex we choose to place it at the origin and write the ansatz in cylidrical coordinates, 
\begin{equation}
\label{psi}
\psi = \sqrt{\mu_\text{b} \over a}\rho\left(r\right)e^{i\phi}\,,
\end{equation} \begin{equation}
\label{A}
\mathbf{A} = {\hbar{q} \over c}{a\left(r\right) \over r} \,\,\boldsymbol{\hat{\phi}} + f(r)\,\,\boldsymbol{\hat{z}}\,,
\end{equation}
where $n_0 = {\mu_\text{b}\over a}$ is the density of the superconductor.  The function $\rho(r)\in (0,1)$ describes the profile function of superconducting density, $\rho(r) = 0$ being no superconducting material and $\rho(r) = 1$ being completely superconducting.  If we assume that the current goes to zero far from the origin we can use \eqref{current} to see that $\lim_{r\rightarrow \infty} a(r) = 1 $ and $\lim_{r\rightarrow \infty} f(r) = 0 $.  The phase of $\psi$ is chosen to mimic a vortex with winding number $n=1$ and only depends on the $\phi$ coordinate.  

In order to calculate long range interactions between vortices we are interested in solutions the equations of motion as $r\rightarrow\infty$.  To decouple our set of differential equations it is convenient to define 
\begin{equation}
\label{sigma}
\rho\left(r\right) = 1 + \sigma\left(r\right)\,,
\end{equation}
\begin{equation}
\label{alpha}
a\left(r\right) = 1 + r\alpha\left(r\right)\,,
\end{equation}
such that $\sigma\left(r\right),\,\,\alpha\left(r\right) \rightarrow 0$ as $r \rightarrow \infty$.  Substituting \eqref{psi} and \eqref{A} into equation \eqref{maxwell} and linearizing yields the two equations,
\begin{equation}
{\partial^{2}\alpha \over \partial{r}^{2}} + {1 \over r}{\partial\alpha \over \partial{r}} - \left({1 \over r^2} + {1 \over \lambda^2}\right)\alpha = 0\,,
\end{equation}
\begin{equation}
\left(\nabla^2  - {1\over \lambda^2}\right)f(r) = {4\pi \over c} j \delta^2(\mathbf{x})\,,
\end{equation}
where $\lambda = \sqrt{\frac{m c^2}{4 \pi q^2 n}}$ is the London penetration depth.
The first equation is the modified Bessel equation of the first order.  We want a solution that goes to zero as $r \rightarrow \infty$ so we choose the solution to be a modified Bessel function of the second kind, $\alpha(r) = {c_\phi \over \lambda}K_{1}\left({r \over \lambda}\right)$. The second equation is just a statement of the Green's function,
\begin{equation}
\label{klein}
(\nabla^2 - \alpha^2)K_{0}(\alpha{r}) = -2\pi\delta^2(\mathbf{x})\,,
\end{equation}
which implies that $f(r) = -{2j \over c}c_\text{z}K_0\left({r\over\lambda}\right)$.
Going back through all the substitutions we find that the vector potential is
\begin{equation}
\label{solArough}
\mathbf{A} = {\hbar c \over q}\left[{1\over r} + {c_\phi\over\lambda}K_{1}\left({r \over \lambda}\right)\right]\,\,\boldsymbol{\hat{\phi}} -{2 j\over c} c_\text{z} K_{0}\left({r \over \lambda}\right)\,\,\boldsymbol{\hat{z}}\,.
\end{equation}
In comparison the standard case without $\mathbf{j}$ we see that a third term has appeared in the interaction.  This attractive component will play the crucial role in what follows. 

A similar procedure follows for the solution to the order field.  Substituting \eqref{psi} and \eqref{alpha} into \eqref{order} and linearizing yields
\begin{equation}
{1 \over r}{\partial \over \partial{r}}\left(r{\partial\sigma \over \partial{r}}\right) = {4m\mu_\text{b} \over \hbar^2}\sigma\,.
\end{equation}
This is a modified Bessel equation of the zeroth order which has a solution, $\sigma(x) = c_{\sigma}K_{0}\left({\sqrt{2} \over \xi}r\right)$.  Substituting this back we find that,  
\begin{equation}
\psi = \sqrt{\mu_\text{b} \over a}\left[1- c_\sigma K_{0}\left({\sqrt{2} \over \xi}r\right)\right]e^{i\phi}\,,
\end{equation}
where $\xi = \sqrt{\hbar^2 \over 2m\mu_\text{b}}$ is the coherence length.

In the next section, when calculating the vortex interactions, it will be useful to ``unwind'' the phase of the vortex.  This will give a much cleaner solution and is done using a gauge transformation \eqref{symmetry}, where $\varphi(\mathbf{x}) = -\left(\hbar c\over q\right)\phi$.  The solutions for the field equations become,
\begin{equation}
\label{solA}
\mathbf{A} = {\hbar c \over q\lambda}c_\phi K_{1}\left({r \over \lambda}\right)\,\,\boldsymbol{\hat{\phi}} -{2 j\over c} c_\text{z} K_{0}\left({r \over \lambda}\right)\,\,\boldsymbol{\hat{z}}\,,
\end{equation} 
\begin{equation}
\label{solsigma}
\psi = \sqrt{\mu_\text{b} \over a}\left[1- c_\sigma K_{0}\left({\sqrt{2} \over \xi}r\right)\right]\,.
\end{equation}
We can now move on to calculating the interactions between vortices.

\section{Interaction between two current carrying vortices}
Superconducting vortices without currents interact through two forces.  There is an attractive force caused by the superconducting order parameter wanting to have one defect instead of two and a repulsive electromagnetic force caused by the charges swirling around the vortex.  Two vortices placed side by side have currents running in opposite directions on their nearest sides and it is well known that opposite currents repel.    

Suppose there are two wires, placed parallel to each other, carrying current.  If the currents run in the same direction the wires will be attracted to one another.  Now consider that superconducting vortices, instead of wires, are carrying the current.  There are three forces working against each other: the attractive electromagnetic force from the current, the repulsive electromagnetic force from the gauge field, and the attractive force from the order field.  If the current were strong enough, the attractive force would be strong enough to completely cancel the force from the gauge field and the vortices would always attract and the superconductor would always act like a type-I.

\subsection{Calculations}
The philosophy behind calculating the interaction between vortices is to find the energy of the entire system and then subtract off the energy of the individual vorticies as originally outlined in \cite{Kramer:1971}.  The technique we will use was introduced in \cite{Speight:1997} and has been used to calculate vortex interactions in models with two order parameters \cite{Buckley:2003zf}, \cite{MacKenzie:2003jp}.  The same philosophy as in \cite{Kramer:1971} is used but the actual calculation becomes much less cumbersome.  We will reduce the theory to a non-interacting, linear one and then model the vorticies as point sources.  The interaction energy is then calculated from this linear theory.

To make the calculation  easier it is useful to use a gauge tranformation to remove the phase in $\psi(\mathbf{x})$.  This is described in the previous section and yields the form $\psi = \sqrt{\mu_\text{b}\over a}(1-\sigma)$.  To linearize the theory we expand in $\rho$ and $\mathbf{A}$ and keep only quadratic terms to get
\begin{eqnarray}
E_\text{free} &=& \int{d^2x}\left\{ {\mu_\text{b} \over a}{\hbar^2 \over 2m}(\nabla\sigma)^2 + 2{\mu_\text{b}^2 \over a}\sigma \right. \nonumber \\ &&  \left. + {1 \over 8\pi}\left((\nabla\times\mathbf{A})^2 + {\mathbf{A}^2 \over \lambda^2}\right) \right\}\,.
\end{eqnarray}
We now add source terms to model the vortices,
\begin{equation}
E_\text{source} = \int{d^2x}\left\{\tau\sigma + \boldsymbol{\mathcal{J}}\cdot\mathbf{A}\right\}\,,
\end{equation}
where $\tau$ and $\boldsymbol{\mathcal{J}}$ are the sources for the fields $\sigma$ and $\mathbf{A}$.  Minimizing this we get the equations of motion,
\begin{equation}
\left(\nabla^2 - {2 \over \xi^2}\right)\sigma = {m \over \hbar^2}{a \over \mu_\text{b}}\tau\,,
\end{equation}
\begin{equation}
\left(\nabla^2 - {1 \over \lambda^2}\right)\mathbf{A} = 4\pi\boldsymbol{\mathcal{J}}\,.
\end{equation}
We want to solve for the sources $\boldsymbol{\mathcal{J}}$ and $\tau$ such that $\sigma$ and $\mathbf{A}$ have the same asymptotic solutions we obtained earlier in \eqref{solA} and \eqref{solsigma}.  Using \eqref{klein}, the derivative of \eqref{klein} with respect to the radial component $r$, and the identity 
\begin{equation}
\label{besselidentity}
{d \over dr}K_0(\alpha r) = -\alpha K_1(\alpha r)\,,
\end{equation}
we can solve for the sources,
\begin{equation}
\label{tau}
\tau = -{\hbar^2 \over m}{\mu_\text{b} \over a}2\pi\delta^2(\mathbf{x})\,,
\end{equation}
\begin{equation}
\label{j}
\boldsymbol{\mathcal{J}} = {\hbar c \over 2q}{\partial\delta^2(\mathbf{x}) \over \partial r}\,\,\boldsymbol{\hat{\phi}} + {j\over c}\delta^2(\mathbf{x})\,\,\boldsymbol{\hat{z}}\,.
\end{equation}

The interaction energy is found by substituting $\boldsymbol{\mathcal{J}} = \boldsymbol{\mathcal{J}}_1 + \boldsymbol{\mathcal{J}}_2$, $\mathbf{A} = \mathbf{A}_1 +
\mathbf{A}_2$, $\tau = \tau_{1} + \tau_{2}$ and $\sigma = \sigma_{1} + \sigma_{2}$ into the total energy $E = E_\text{free} + E_\text{source}$ and subtracting of the energies of the vortices, leaving only cross terms.  The subscripts $1$ and $2$ refer to two separate vortices and positions $\mathbf{x}_1$ and $\mathbf{x}_2$ respectively.  Using the equations of motion we get left over cross terms that are interpreted as the interaction energy;
\begin{equation}
E_\text{int} = \int{d^2x}\left\{ \tau_{1}\sigma_{2} + \boldsymbol{\mathcal{J}}_1\cdot\mathbf{A}_2\right\}\,.
\end{equation}
Though it is not apparent, the interaction energy is symmetric in the exchange of the subscripts $1$ and $2$.  The apparent asymmetry arises when the equations of motion for either subscript $1$ or $2$ are substituted in.  Using \eqref{A}, \eqref{sigma}, \eqref{tau} and \eqref{j} the interaction energy can be written\footnote{We are thankful to Maxim Lyutikov who pointed out the missing factor "c" in the expressions (\ref{current_interaction},  \ref{currentinequality}).},
\begin{eqnarray}
E_\text{int} &=& \int{d^2x}\left\{ -{\hbar^2 \over m}{\mu_\text{b} \over a}2\pi\delta^2(|\mathbf{x-x}_1|)K_{0}\left({\sqrt{2} \over \xi}|\mathbf{x-x}_2|\right) \right. \nonumber \\ &&  \left. - {\hbar^2 c^2 \over 2 q^2 \lambda}{\partial\delta^2(\mathbf{x-x}_1) \over \partial r}K_{1}\left({|\mathbf{x-x}_2| \over \lambda}\right) \right. \nonumber \\ &&  \left. - {2 j_1 j_2 \over c^2} \delta^2(\mathbf{x-x}_1) K_{0}\left({|\mathbf{x-x}_2| \over \lambda}\right)  \right\}\,,   \nonumber \\
&=& {1\over2}\left({\hbar c\over q \lambda}\right)^2\left[\left(1- {4q^2 \lambda^2 j_1 j_2 \over \hbar^2 c^4}\right)K_{0}\left({d \over \lambda}\right) \right. \nonumber \\ &&  \left. - K_{0}\left({\sqrt{2}d \over \xi}\right)\right] \label{current_interaction},\,
\end{eqnarray}
where $d = |\mathbf{x}_1 - \mathbf{x}_2|$.  To evaluate the second term in the integral we made use of equations (\ref{besselidentity}) and (\ref{klein}).  

If $j_1$ and $j_2$ are set to zero we obtain the interaction between gauge vortices without current.  The only new piece in the interaction is that which comes directly from the current.  If $j_1$ and $j_2$ run in the same direction there is an attractive force and if they run in opposite directions there is a repulsive force.  This is the expected result if we considered parallel wires carrying current.

The interaction energy \eqref{current_interaction} determines whether the vortices attract or repel and whether we see type-I or type-II behaviour in the superconductor.  If we set $j_1 = j_2 = j$ then there are two cases to explore; one when $j<{\hbar c^2 \over 2 q \lambda}$ and one when $j>{\hbar c^2 \over 2 q \lambda}$.  In the first case the first term of \eqref{current_interaction} is positive and we obtain the canonical behaviour for a superconductor where the Landau-Ginzburg parameter decides whether the system exhibits type-I or type-II behaviour.

\subsection{Discussions}
First, let us consider the regime when current is large,
\begin{equation}
\label{currentinequality}
j > {\hbar c^2 \over 2 q \lambda}\,,
\end{equation}
and the first term in \eqref{current_interaction} becomes negative.  This means that there is no longer a repulsive term present in the interaction and the vortices will always attract.  This is the main result of this paper.  If the condition \eqref{currentinequality} is met, all the components of the interaction become attractive.  While a naive calculation of the Landau-Ginzburg parameter suggests it is a type-II superconductor, it actually behaves like type-I, and the conflict is resolved.  

In this case an intermediate state will be formed, suggesting that alternating domains of superconducting and normal matter coexist in this regime. This state would be  the lowest energy state on the phase diagram with given $B, j$  satisfying condition (\ref{currentinequality}). This implies that type-II vortices don't form at any time in the neutron star's life, even during the short period of cooling when the transition to a superconducting state takes place.  

One should remark that due to the static nature of the problem the result \eqref{currentinequality} persists for relativistic systems, which may be relevant for study color super conducting phases.  The equations of motion used to derive \eqref{currentinequality} remain unchanged if we replace $\sqrt{\hbar^2 \over 2m}\psi \rightarrow \phi_\text{rel}$ and redefine the corresponding coupling constants.  

Now let us consider a more realistic case when the current is small,
\begin{equation}
\label{currentinequality1}
j < {\hbar c^2 \over 2 q \lambda}\,.
\end{equation}
In this case we cannot make any precise statements within our framework.  However, based on experience in similar situations in condensed matter systems one should expect very dramatic changes to the vortex lattice when there are currents directed along the external magnetic field \cite{Clem:1977},\cite{Brandt:1982},\cite{Marsh:1994},\cite{Genenko:1995},\cite{Kohandel:2000},\cite{Saveliev:2003}.  

In this literature, the presence of longitudinal currents is shown to cause a vortex to develop a spiral-vortex instability.  This instability can be delayed for small currents or even stabilized due to the impurities. The lesson from these condensed matter systems is that when a current aligns with the magnetic field of the vortex the properties of the vortex lattice are completely changed or destroyed.
 
We expect similar behaviour in regions of the neutron star where both the Landau-Ginzburg parameter suggests type-II behaviour and longitudinal currents are induced.  While many features of the system are still to be explored, the main massage for the present study that it is very likely (similarly to CM studies mentioned above) that even small currents (\ref{AV}) can completely destroy the vortex lattice by replacing it with a new still unknown structure.  

It is not our purpose to discuss the rich physics related to vortex instabilities resulting from longitudinal currents, but rather stress that the resulting state will definitely be not the rigid Abrikosov lattice.  It is unclear what structure will replace the Abrikosov lattice but it is reasonable to believe that superconductivity would persist in this new regime; the energy scales associated with currents are much smaller than the superconducting gap. 

It is possible (but not necessary) that the intermediate state typical for type-I superconductivity will develop and alternating domains of superconducting and normal matter would coexist. The size and shape of the domains are known to be very sensitive to many things: geometry, initial conditions, the method of preparation of a sample, boundary conditions,  surface effects.  As is known, the intermediate state is not in thermodynamic equilibrium in the strict thermodynamical sense, but rather depends on the history of the system. It is also possible that other states, such as Bragg glass phase \cite{Kohandel:2000} would develop, or vortex -lattice melting transition would take place \cite{Saveliev:2003}. 

The exact state is not essential at the moment.  What is essential is that the Abrikosov lattice is destroyed by longitudinal currents. There are many alternative states that may replace the Abrikosov lattice.  We shall refer to the absence of the Abrikosov lattice (which is a consequence of  type-II superconductivity) as a type-I superconductor that supports the intermediate state even though many other phases may result from current induced vortex instabilities. Therefore, the conflict between the precession of a neutron star and the standard estimation of the  Landau-Ginzburg parameter likely will be resolved even when induced currents are small.

\section{Conclusion and Speculations}
If currents are induced in vortices then we have found a mechanism that reconciles the condradiction between the precession of neutron stars and the standard presumption that there is type-II superconductivity inside a neutron star.  A sufficiently strong current running along the core of the vortex and satisfying inequality \eqref{currentinequality} allows the vortices to attract even if the Landau-Ginzburg parameter indicates they should repel.  

A neutron star would rather form the domain structure seen in type-I superconductors rather than the vortex lattice structure seen in type-II superconductors, thus resolving the puzzle. We also argued that even small currents along magnetic field can completely change/destroy the structure of the Abrikosov lattice. 

A pertinent question is whether these currents can actually be induced in neutron stars.  The answer depends crucially on the details of the specific phase realized in the core of a neutron star. As formulated in section III, the electromagnetic currents will be induced if the Goldsone modes condense in the presence of a background magnetic field.  

If we assume this is the case then there are many questions to be considered.  How would the magnetic field be distributed?  What is the fate of these currents? 

{\bf A)} If the current is large, it is expected that the magnetic field could exist in macroscopically large regions where there are alternating domains of superconducting (type-I) matter and normal matter - the so called intermediate state.  It has been estimated long ago \cite{Baym} that  it takes a very long time to expel a typical magnetic flux from the neutron star core. Therefore, if the magnetic field existed before the neutron star became a type-I superconductor (before it sufficiently cooled down), it is likely that the magnetic field will remain there.  

The intermediate state is characterized by alternating domains of superconducting and normal matter where the superconducting domains exhibit the Meissner effect, while the normal domains carry the required magnetic flux.  The pattern of these domains is strongly related to the geometry of the problem. The simplest geometry, originally considered by Landau \cite{LandauLam}, is a laminar structure of alternating superconducting and normal layers.  While precise calculations are required for understanding of the magnetic structure in this case \footnote{ It is clear that the corresponding calculations would require an understanding of the non-equilibrium dynamics.  Indeed, the resulting picture of the system would be very different if magnetic field is turned on before the system becomes superconducting or after. This unambiguously implies that the corresponding state is not in thermodynamic equilibrium.}, one can give some simple  estimation of  the size of the domains using the calculations Landau presented for a different geometry.  His formula \cite{LandauLam} suggests that the typical size of a domain is
\beq
\label{domain}
a\sim 10 \sqrt{R \Delta}\,,
\eeq
where $R$ is a typical external size identified with a neutron star core ($R\sim 10$ km), while $\Delta$ is the typical width of the domain wall separating normal and superconducting states.  We estimate $\Delta\sim \lambda $ as the largest microscopical scale of the problem. Numerically, $a\sim 10^{-1}$ cm which implies that a typical domain can accommodate $\sim 10^4$ neutron vortices separated by a distance $\sim 10^{-3}$ cm.
 
{\bf B)} What is the fate of these currents?  It is quite possible that the current will travel inside the superconducting region in only one direction. The current is conserved, so it must make a ${\bf U}$-turn and start travelling in the opposite direction either along the crust of the neutron star or through regions of normal matter in the intermediate state.  This is similar to the   case of $^3$He-A system 
discussed in ref.\cite{volovik1} (see Fig.23 from that reference).  

{\bf C)} If the current on the way back travels through the normal matter of  a different domain (or the crust) then large current loops with typical sizes comparable to the neutron star radius would be created.  These large current loops would induce a coherent toroidal magnetic field that, when combined with the poloidal field present in a neutron star, would create a non-zero magnetic helicity.  

Such a toroidal magnetic field is apparently necessary to describe the temperature distribution of the crust \cite{Page:2005}. It has been also argued long ago that a toroidal component in the magnetic field of a neutron star is necessary for stablility of the poloidal magnetic field \cite{Markley:1973}.  

{\bf D)} The existence of a current making a ${\bf U}$-turn near the surface of the star may be a key to the understanding of the long standing problem of neutron star kicks \cite{kick,kick1}.  As is known, pulsars exhibit rapid proper motion characterized by a mean birth velocity of $450 \pm 90$ km/s.  Their velocities range from 100 to 1600 km/s \cite{kick} with about 15$\% $ of all pulsars having speeds over 1000 km/s\cite{kick1}.  Pulsars are born in supernova explosions so common theories naturally to look for an explanation in the internal dynamics of the supernova.  However, three-dimensional numerical simulations \cite{kick2} show that even the most extreme, asymmetric explosions do not produce pulsar velocities greater than 200 km/s. Therefore, a different explanation is needed.  

The origin of these velocities has been the subject of intense study.  Many of the theories involve an assymetry in the star's structure, and indeed, many mechanisims are capable ``in principle" of producing the required asymmetry.  In the presence of an external magnetic field, the neutrinos produced in the star are automatically asymmetric with respect to the direction of $\vec{B}$.  However, the most common problem with the suggested mechanisms is the difficulty of delivering the produced asymmetry to the surface of the star. Only when the assymetry reaches the surface of the star may it result in producing the proper motion of the entire star.  

Due to their topological nature, the current (\ref{AV}) may be capable of delivering the required asymmetry produced in the interior of the star to the surface without dissipation.  Even in a strongly interacting theory, the current (\ref{AV}) is persistent and non-dissipating.  In an environment as unfriendly as the dense quark/nuclear matter in neutron stars there is still no dissipation due to re-scattering, and can be effectively used to deliver information across the bulk of the star.  

When the current makes a ${\bf U}$-turn on the surface, a large amount of momentum (due to photon emission) can be transfered to the star.  Therefore, this is a unique opportunity to use our topological currents (\ref{AV}) for delivering the asymmetry produced in the bulk of the star (e.g. due to the Goldstone condensation) to solve the problem of neutron star kicks \cite{kick,kick1}.  

One should notice that the currents may not satisfy constraint  eq. (\ref{currentinequality})  for the explanation of neutron star kicks.  Indeed, relatively small  current is still capable of transfering momentum because the ${\bf U}$-turn mechanism remains operative.

This asymmetry mechanism is different from most in that it does not occur during the supernova, but over a long period of time.  The momentum transfered from each emmited photon is small but if given long enough is sufficient to accelerate the neutron star to the observed proper velocities.    

{\bf E)} A different, but likely related phenomena, is the recent observation of pulsar jets \cite{Pavlov:2000} which are apparently related to neutron star kicks \cite{Wang:2005jg, Lai:2000pk}.  It has been argued that spin axes and proper motion directions of the Crab and Vela pulsars are aligned.  Such a correlation would follow naturally if we suppose that the kick is caused by a non-dissipating current, as suggested in {\bf D)}.  The current, and thus the proper motion, is aligned with the magnetic field, which itself is correlated with the axis of rotation.  

As we mentioned above, the ${\bf U}$-turn mechanism is necessarily accompanied by the photon emission (which delivers the momenta required for the neutron star kick). It would be very tempting to identify the observed inner jets \cite{Pavlov:2000} with the photons emitted when the current makes the ${\bf U}$-turn and starts travelling in the opposite direction along the crust.  In this sense the mechanism for the kick is similar to the electromagnetic rocket effect suggested previously\cite{Lai:2000pk}.
 
{\bf F)} What else could happen with vortices near the crust?  
Reference \cite{Melatos:2005} presents calculations in which vortices near the crust are grabbed and bundled together by the Kelvin-Helmoltz [KH] waves created by the instability that arises when there is shear stress between two fluids. This bundling of vortices could twist them so they no longer line up in an array, but instead form vortex loops (vortons). These vortex loops lie in a plane perpendicular to the angular momentum rather than than along it.  

Such surface KH instability may explain pulsar glitches\cite{Mastrano:2005xw}.  Spiral vortex instability \cite{Clem:1977} observed in condensed matter systems may also have some relation to the formation of these vortons and to glitches.  The helical structure of the vortex could expand to the surface and transfer its angular momentum to the crust. 

The vortex loops made from superfluid vortices are not typically stable (similar to cosmic strings \cite{Witten:eb}) but the presence of a current in the core of a superfluid vortex and an external magnetic field could make these loops stable.  If vortons are stable then they could attract to each other and form a column.  Since they attract very close to each other this structure would cease to look like a bunch of individual vortices but like a single cylindrical vortex sheet which carries a surface current similar to a solenoid.  Vortex sheets have been studied in the context of superfluidity by Landau and Lifshitz \cite{Landau:1955}.

These and many other consequences of this picture still remain to be explored.

\begin{acknowledgments}
The authors would like to thank Jeremy Heyl for useful discussions and Lars Bildsten for the discussions during his visit to Vancouver.  The authors would also like to thank Bennett Link, Maxim Lyutikov and 
Armen Sedrakian   for their many useful comments.  This work was supported in part by the National Science and Engineering Research Council of Canada. 
\end{acknowledgments}

\end{document}